\documentclass[a4paper]{article}
\usepackage{graphicx}

\begin{document}

\begin{center}
{\Large \bf Toward simulation of quark and diquark fragmentations in the
            Quark-Gluon String Model (QGSM) }
\end{center}

\begin{center}{
 V. Uzhinsky\footnote{Laboratory of Information Technologies, JINR, Dubna, Russia.},
 A. Galoyan\footnote{Veksler and Baldin Laboratory of High Energy Physics, JINR, Dubna, Russia.}
}\\
on behalf of Geant4 hadronic working group
\end{center}

\begin{center}
\begin{minipage}{14cm}
Within the Quark-Gluon String Model A.B. Kaidalov found a behaviour of quark and diquark fragmentation functions
for $z\rightarrow 0$ and $z\rightarrow 1$, and proposed interpolation formulae for the functions in the whole
region of $z$. These functions must be a solution of the well-known system of the integral equations. A simplified
Monte Carlo estimation of the functions, based on usage of the fragmentation functions at $z\rightarrow 1$ as the
kernel functions of the system, does not reproduce Kaidalov's results. An improvement of the Monte Carlo
simulations is proposed in this paper. It can be implemented in Monte Carlo event generators such as Los Alamos
QGSM, QGSJet-II and the Geant4 QGS model. It will improve a description of experimental data in the models,
especially, the description of the latest NA61/SHINE Collaboration data on $\pi{\rm C}$ interactions.
Description of the data is a problem in DPMJet, QGSJet, EPOS and Sibyll models.
\end{minipage}
\end{center}

The Quark-Gluon String Model (QGSM) is based on the dual topological unitarisation, the reggeon phenomenology and
parton ideas. It is aimed on description of soft particle production in hadron-hadron, hadron-nucleus and
nucleus-nucleus interactions. It is now implemented in the well-known event generators -- the Los Alamos QGSM
\cite{LAQGSMa,LAQGSMb}, the Shield code \cite{Shield}, the Geant4 QGSM \cite{G4QGSMa,G4QGSMb} and QGSJet
\cite{QGSJet}.

The model was proposed in the papers \cite{QGSM}. In particular, the quark and diquark favored fragmentation
functions in QGSM were found in Kaidalov's papers \cite{FFkaidalov} with the help of 3-reggeon phenomenology:
\begin{eqnarray}
D_Q^h(z)=\frac{c_h}{z}, ~~~ ~~~ ~~~ ~~~ ~~~ ~~~ ~~~ ~~~ ~~~ ~~~ ~~~ ~~~ ~~~ ~~~ ~~ z\rightarrow 0; \label{Eq1}\\
D_Q^h(z,p_T^2)=G_Q^h(p_T^2)\ (1-z)^{\alpha_{i\bar i}(0)-2\alpha_{i\bar k}(-p_T^2)}, ~~~ z\rightarrow 1,
\label{Eq2}
\end{eqnarray}
where $Q$ is an identificator of a quark system -- $u$, $d$, $s$, $\bar u$, $\bar d$, $\bar s$, $uu$, $ud$ etc.;
$h$ represents a hadron -- $\pi$, $\rho$, $K$, $p$, $\bar p$ etc. $z$ is a momentum fraction of
quark system carried by a produced hadron $h$ in the infinite momentum frame. $p_T$ is transverse momentum of
hadron $h$ relative to the original quark system momentum. $G_Q^h(p_T^2)$ is a distribution of a hadron on $p_T$.
$c_h$ is a constant which depends on type of the produced hadron $h$. $\alpha_{i\bar i}(0)$ is an intercept of the
reggeon trajectory with quark content $i\bar i$. $\alpha_{i\bar k}(-p_T^2)$ is the reggeon trajectory on which
hadron $h$ with quark content $i\bar k$ is located ($\alpha_{i\bar k}(-p_T^2)=\alpha_{i\bar k}(0)+\alpha_{i\bar
k}'\cdot p_T^2$).
In the case of unfavored fragmentation, $D_Q^h(z,p_T^2)$ is multiplied on $(1-z)$.

Interpolation formulae for the fragmentation functions valid in the whole interval of $z$ ([0,1]) were also given.
Some of them are shown below\footnote{The author of Refs.~\cite{FFkaidalov} assumed that fragmentation functions of
$u$, $d$ quarks into resonances (for example $\rho$) are also given by Eqs. \ref{Eq3} and \ref{Eq4} with
substitution $c_\pi \rightarrow c_\rho$.}
\begin{eqnarray}
D_u^{\pi^+}=D_d^{\pi^-}&=&\frac{c_{\pi}}{z} (1-z)^{-\alpha_\rho+\lambda},    \label{Eq3}\\
D_u^{\pi^-}=D_d^{\pi^+}&=&\frac{c_{\pi}}{z} (1-z)^{-\alpha_\rho+\lambda+1},  \label{Eq4}\\
D_{uu}^{\pi^+}&=&\frac{c_{\pi}}{z} (1-z)^{\alpha_\rho-2\alpha_N(0)+\lambda}, \label{Eq5}\\
D_{ud}^{p}&=&\frac{1}{z} \left[ C_N z^{2(\alpha_\rho(0)-\alpha_N(0))} (1-z)^{\alpha_\rho+\lambda}\right.\label{Eq6} \\
& &+ \left. c_N (1-z)^{-\alpha_\rho(0)+\lambda+4(1-\alpha_N(0))}\right],\nonumber
\end{eqnarray}
$$\lambda=2\alpha_\rho'p_T^2.$$


As known \cite{SEquA,SEquB,SEquC}, the fragmentation functions are a solution of the integral equations:
\begin{equation}
D_Q^h(z)=f_{Q,Q-\bar h}^h(z) + \sum_{h'} \int^1_z \frac{dz'}{z'}\ f_{Q,Q-\bar h'}^{h'}(1-z')\ D_{Q-\bar
h'}^h(\frac{z}{z'}), \label{Eq7}
\end{equation}
where $f_{Q,Q-\bar h}^h$ are the kernel functions, and $\bar h$ is the antiparticle corresponding to the hadron
$h$; $f_{Q,Q-\bar h}^h(z) dz$ is the probability that the system $Q$ emits the hadron $h$ with momentum fraction from $z$
to $z+dz$.

The kernel functions needed for a Monte Carlo simulation of quark and diquark fragmentations are not known in
QGSM. For $z\rightarrow 1$ $D_Q^h(z)\simeq f_{Q,Q-\bar h}^h$. Thus, in practice $z\ D^h_Q$ given by Eqs. \ref{Eq3}
-- \ref{Eq6} and others are used as kernel functions. Additional to this, $\lambda \rightarrow
2\alpha_\rho'<p_T^2> \sim$ 0.5. In general, it is not ensure than a solution of Eq. \ref{Eq7}
with these simplifications will
give the initial $D^h_Q$. For example, this happens in the Los Alamos QGSM \cite{LAQGSMa,LAQGSMb}, Shield
\cite{Shield} (see function {\it ZFRAGS}) and in QGSJet-II \cite{QGSJet} (see subroutine {\it qggene},
$\lambda\simeq 1.6$). The same happened with Geant4 QGSM before 2015. Results of these simplifications for
$\alpha_\rho=0.5$ and $\lambda = 0.5$ are shown in Fig.~1 by dashed lines.
\begin{figure}[bth]
\begin{center}
\includegraphics[width=140mm,height=50mm,clip]{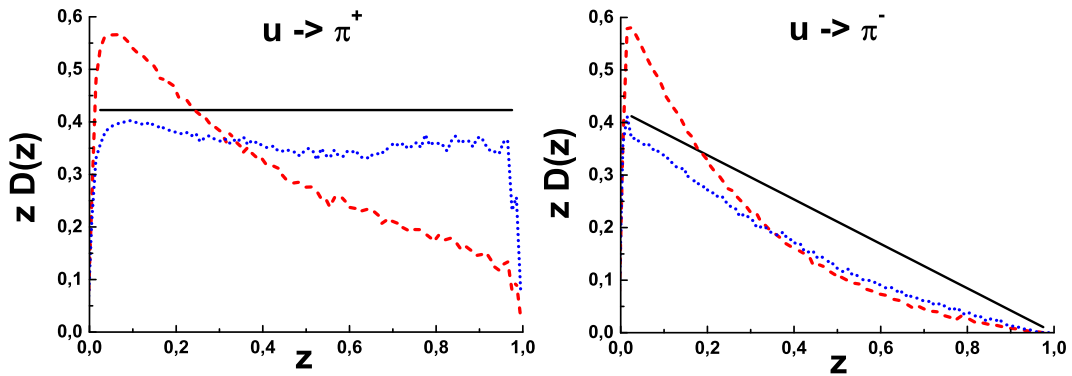}
\end{center}
\caption{Functions $zD_Q^h(z)$ for u-quark fragmentation into $\pi^+$ and $\pi^-$ mesons.
         Solid lines are functions given by Eqs. 3 and 4 for $\alpha_\rho=0.5$ and
         $\lambda = 0.5$. Dashed lines are Monte Carlo results of the simplification
         described above.
         Dotted lines are the simulation results for $b=2$ (see text below).}
\label{Fig1}
\end{figure}

As seen, the Monte Carlo estimation of $zD_Q^h(z)$ for $u\rightarrow \pi^+$, as well as for $u\rightarrow \pi^-$,
is finite for both $z\rightarrow 1$ and $z\sim 0$. The decreasing of $zD_Q^h(z)$ for growing $z$ is
connected with mesonic resonance decays. $zD_Q^h(z)$ is a constant on the level $\sim$ 0.12 for directly produced
$\pi^+$ mesons (see dashed line at $z\sim$ 0.95). The last value is quite understandable: a $u$ quark can couple
with a see $\bar d$ antiquark with a probability $\sim 0.415$. A created meson can be a $\pi^+$ meson with
a probability $\sim 0.5$, or can be a $\rho^+$ meson with a probability $\sim 0.5$. Thus, a
probability of the process ($u\rightarrow \pi^+$) is approximately equal $0.415\cdot 0.5\simeq 0.2$. The probability
would decrease if one takes into account the diquark-antidiquark pair production from the vacuum.

To improve the behaviour of the estimations, let us assume that the kernel functions can have
a multiplier $z^b$ for $z\rightarrow 1$ ($f_{Q,Q-\bar h}^h(z)=c_h (1-z)^a z^b$). The tuning of $b$ gave
results presented by dotted lines in Fig.~1 for $b=2$. As seen, the simulation results become now
close to the Kaidalov's approximations. At least, the forms of the functions coincide.

Analogous changes can be done for other kernel functions, $u\rightarrow \rho^+/\rho^0$,
$u\rightarrow K^+/K^0$, $u\rightarrow p$, $s\rightarrow K^-/K^0$ and so on. Various values
of $b$ have to be used for various processes. In particular, we used $b=2$ for $u/d\rightarrow \rho^0$.

A direct consequence of the improvement can be seen by comparing the simulation with actual experimental data.
Recently, the NA61/SHINE Collaboration has published experimental data on $\rho^0$ meson
production in $\pi^-{\rm C}$ interactions at $P\-{lab}=158$ and 350 GeV/c \cite{NA61rho}.
The authors of the paper have compared their data with predictions of various Monte Carlo event
generators. Some of the predictions are shown in Fig.~2 together with Geant4 QGS model calculations
for $\alpha_\rho=0.5$, $\lambda = 0.5$ and $b=2$.

As seen, the improved Geant4 QGS model nicely reproduces the data on $\rho^0$ meson
production at $x_F\geq 0.3$. Without the improvement, the results were very close to the Geant4
FTF model calculations (see long dashed curve in the figure). The QGSJet model \cite{QGSJet}, also
based on QGS model does not reproduce the shape of $\rho^0$ meson distribution. This is strange because
at $\lambda = 1.6$ the $\rho^0$ meson distribution must be a decreasing function of $x_F$. There
might be in the QGSJet code a special treatment of $\rho$ meson production.

The EPOS model \cite{EPOS}, as known, uses QGS model ideas, but it uses fragmentation functions
tuned for $e^+e^-$ annihilations. Thus, the distributions of $\rho$ and $\omega$ meson must be
decreasing functions of $x_F$. This is true for $\omega$ meson production, but it is not so for
$\rho^0$ meson distributions (see red short-dashed lines in Fig.~2). We believe this is also
connected with a special treatment of $\rho$ meson production. In contrast, the Geant4 QGS model
predicts similar shapes for $\rho^0$ and $\omega$ mesons' distributions.

The Geant4 FTF model is a modern implementation of the well-known Fritiof model \cite{Fri1,Fri2}. It uses the LUND
algorithm for fragmentation of strings \cite{SEquB}. The corresponding fragmentation functions were tuned to
describe $e^+e^-$ annihilations. As seen in Fig.~2, these lead to distributions of $\rho$
and $\omega$ mesons decreasing with $x_F$.
\begin{figure}[bth]
\begin{center}
\includegraphics[width=140mm,height=90mm,clip]{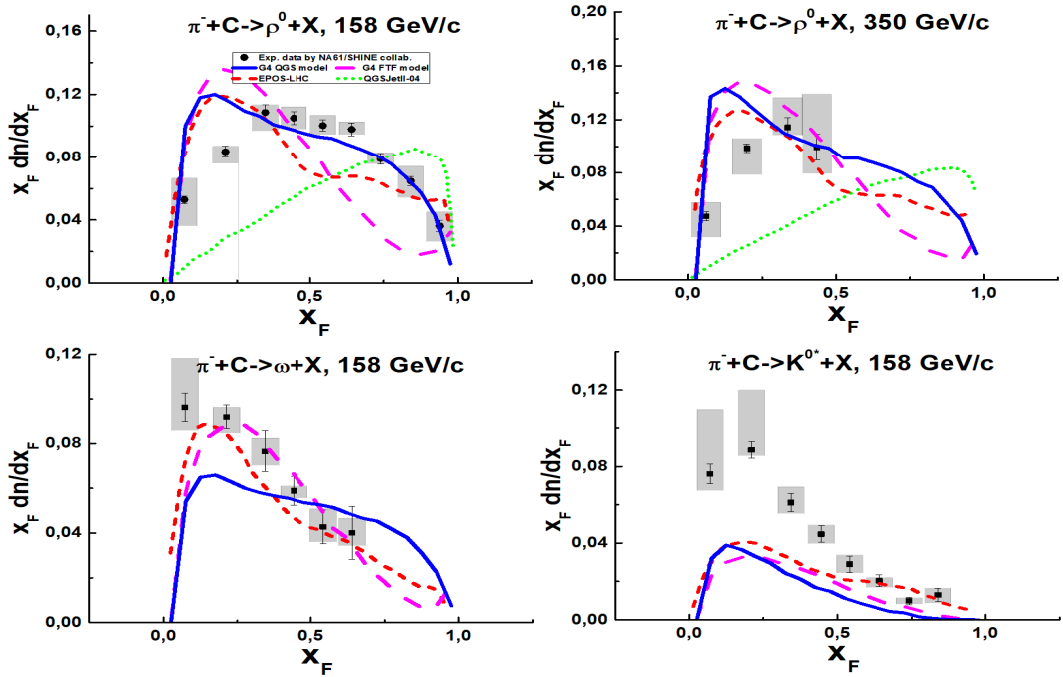}
\end{center}
\caption{Mesonic resonances distributions as functions of $x_F$. Points with statistical and systematical
         errors are experimental data \protect{\cite{NA61rho}}. Solid (blue) lines are calculations of
         the improved Geant4 QGS model. The long-dashed (purple) lines are results of the Geant4 FTF
         model simulations. The short-dashed (red) and dotted (green) lines are calculations by the EPOS and QGSJet
         models, respectively.} 
\label{Fig2}
\end{figure}

A summed yield of $\rho^0$ and $\omega$ mesons in the central region ($x_F\sim$ 0) depends on the probability of
vector meson production. Usually, it is assumed in Monte Carlo models that probabilities of pseudo-scalar and
vector meson productions are related as 0.5:0.5. To describe the yield in the Geant4 QGSM, we assume that the
ratio is 0.25:0.75. A relation between $\rho^0$ and $\omega$ mesons depends of the mixing of these mesons. A
commonly accepted mixing is 0.5:0.5 ($\rho^0$ and $\omega$ mesons are produced with equal probabilities). The data
require the mixing 0.625:0.375 (a dominant production of $\rho^0$ mesons). We have to note that the leading
particle effect also has an influence on these meson distributions at $x_F \sim$ 1.

A more complicated situation takes place with the $K^{0*}$ meson production. As seen in Fig.~2, all the models do
not describe $K^{0*}$ meson distribution -- the predictions are essentially below the experimental data for small
$x_F$. If we allow only vector meson production in the Geant4 QGSM, it will increase the yield of $K^{0*}$ mesons
only on 32 \%! It is not enough.

Another possibility for enhancing the $K^{0*}$ meson production is to increase the $s\bar s$ pair production from
the vacuum during string fragmentations. It is assumed in the standard version of the Geant4 QGS model that this
probability is equal to 17 \%. Setting probabilities of $u\bar u$, $d\bar d$ and $s\bar s$ pair productions from
the vacuum as 1/3:1/3:1/3, it is possible to increase the $K^{0*}$ meson production, and reach the aimed
description of experimental data. Unfortunately, in this case, multiplicities of $K^+$ and $K^-$ mesons are
strongly overestimated.

In order to understand a source of the disagreement between the data and the model calculations, let us reapply
in part the method of resonance yield extraction used in Ref.~\cite{NA61rho}. In this paper, the collaboration studied the
effective masses of positive and negative charged particle pairs in $\pi^-{\rm C}$ interactions. To each charged
particle the $\pi$-meson mass was assigned. After that, effective masses of the pairs were calculated, and primary
distributions on the pair mass in various intervals of pair $x_F$ were obtained. The mass distribution of pairs
with the same charge was considered as a background. For an extraction of resonance yields, the "templates"
technique was used. A template is a mass distribution of "$\pi^+$" and "$\pi^-$" pairs produced by the decays of a
selected resonance (see Fig.~15 in \cite{NA61rho}). A primary distribution was considered as the sum of the
background and the templates with some fitting coefficients. The integral of a template with a corresponding
coefficient normalized on interaction event number was defined as a corresponding resonance yield.

We show in Fig.~3 opposite-charge mass distribution of pairs with $x_F>$0 with subtracted background for our
simulated events. The distribution shows a peak at $M_{eff}\sim$ 800 MeV associated with $\rho^0$ meson
production, a plateau at $M_{eff}\sim$ 450 -- 550 MeV connected with $\omega$ mesons, and a peak at $M_{eff}\sim$
350 MeV due to decays of $\eta$ mesons. A peak connected with $K^{0*}$ cannot be directly seen in this
distribution. To show the $K^{0*}$ peak clearly, we plot in Fig.~3 by dashed line the pair mass distribution in
events simulated with forbidden $\rho^0$ decays. An analogous distribution was obtained by disabling  the
simulation of all resonance decays (see dotted line in Fig.~3). The last distribution is formed due to directly
produced light mesons. As seen, the peaks are still presented in the distribution.
\begin{figure}[bth]
\begin{center}
\includegraphics[width=140mm,height=100mm,clip]{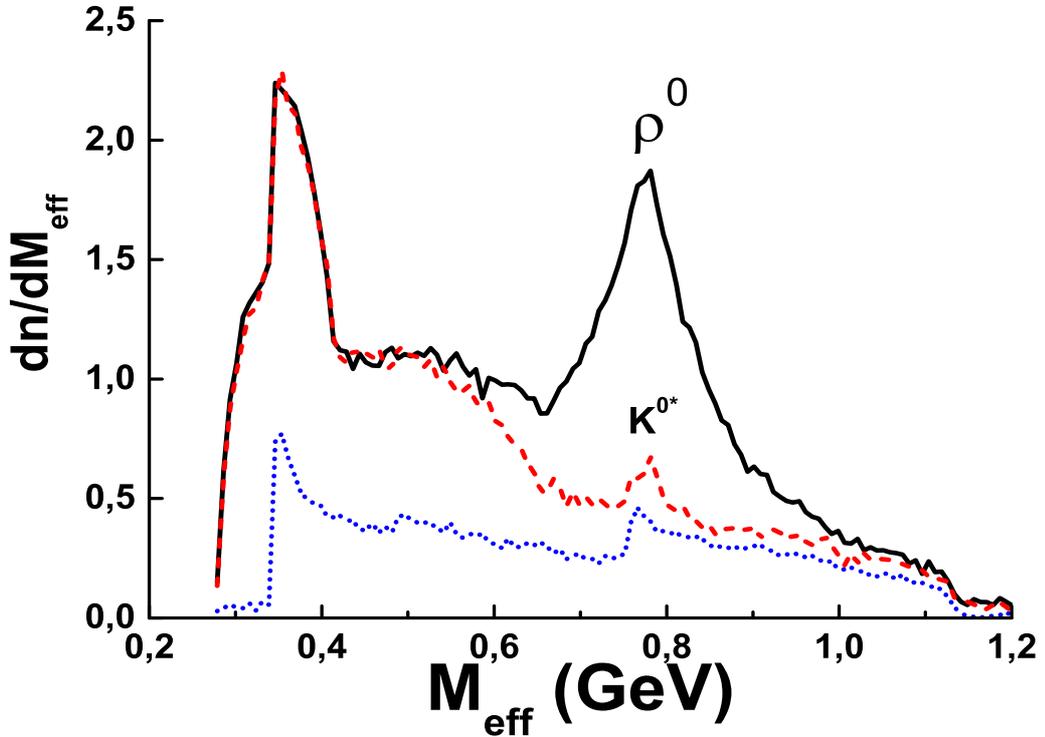}
\end{center}
\caption{Opposite-charge pair mass distributions in $\pi^-{\rm C}$ interactions at 158 GeV/c without background.
         The solid (black) line presents calculations with all allowed resonance decays. The dashed (red) line shows
         calculations with forbidden $\rho^0$ decays. The dotted (blue) line is a distribution of directly produced
         light meson pair masses.}
\label{Fig3}
\end{figure}

The peaks in the directly produced meson mass distribution are connected with decay of strings with small masses.
It is assumed in all Monte Carlo generators that a fragmentaion of a string stops when a mass of the string
reaches a sufficiently small value. We assume in the Geant4 QGS model that this value is equal to 350 MeV for
$u-\bar u$ and $d-\bar d$ strings, and 710 MeV for $s-\bar d$ and $\bar s-d$ strings\footnote{The strings decay
into $\pi$ and $K$ mesons. Due to the redefinition of $K$ meson masses, the corresponding peak position is shifted
to higher values.}. Of course, the values can be smeared, as it is doing in the Pythia code. But a presence of the
small mass string peaks in the pair mass distributions is an open question.

Returning to the $K^{0*}$ production, we note that the estimation of $K^{0*}$ yields from the pair mass
distributions without considering of the direct meson distributions (see dashed and dotted lines in Fig.~3) can
lead to an overestimation of the yield approximately a factor of two. By decreasing the corresponding experimental
data presented in Fig.~2 by a factor of two, we can reach an agreement with the model calculations. We believe
that the model calculations were done without applying the experimental method of the resonance yield
estimation. A reproduction of the method requires a Geant4 description of the experimental setup, energy
depositions of particles and responses of detector systems, reconstruction and identification of tracks, and
many other things. It can be done only by the NA61/SHINE collaboration. We expect that this correct application
of the method will erase the large disagreement between the experiment and the theory
on yield of $K^{0*}$ meson productiion.

\section*{Conclusion}
\begin{enumerate}
\item As shown above, it is possible to reproduce Kaidalov's fragmentation functions of quarks and diquarks into
hadrons in a Monte Carlo approach;

\item Implementation of the proposed improvement in the Geant4 QGS model allowed us to describe the latest data
of the NA61/SHINE Collaboration on $\rho$ and $\omega$ meson production in $\pi^-{\rm C}$ interactions;

\item The description of the data requires probabilities of pseudo-scalar and vector meson productions as 0.25:0.75,
the mixing of $\rho$ and $\omega$ mesons as 0.625:0.375, and probabilities of $u\bar u$, $d\bar d$ and $s\bar s$
pair productions as 0.415:0.415:0.17;

\item Extracted yield of $K^{0*}$ mesons can depend on the yield of the small mass string fragmentation.
A correct comparison of Monte Carlo simulation results with the experimental data requires to take into account
the detailed experimental analysis into the simulations.

\end{enumerate}

The authors are thankful to the Geant4 hadronic working group, especially, G. Folger and A.  Ribon for
useful consideration and interest to the work, and SFT division of CERN for a financial support.

\end{document}